\documentclass[amsmath,amssymb,
showpacs,twocolumn,
superscriptaddress,
prl]{revtex4-1}
\usepackage{graphicx,bm,color,subfigure}
\usepackage[T1]{fontenc}
\usepackage{mathrsfs}
\usepackage{bm}
\usepackage{amsmath}
\usepackage{amssymb}
\usepackage{graphicx}
\usepackage{esint}
\usepackage{multirow}
\usepackage{float}
\usepackage{array}
\usepackage{makecell}
\usepackage{harpoon}
\usepackage{booktabs}
\usepackage{gensymb}
\usepackage{simplewick}
\usepackage{subfigure}
\usepackage{soul}

\def\Eq#1{Eq.~(\ref{#1})}
\def\Fig#1{Fig.~\ref{#1}}

\definecolor{ZXBlue}{RGB}{0,125,255}

\newcommand{\up}{\uparrow}
\newcommand{\dn}{\downarrow}

\newcommand{\be}{\begin{eqnarray}}
\newcommand{\ee}{\end{eqnarray}}
\newcommand{\la}{\langle}
\newcommand{\ra}{\rangle}



\begin{document}

\title{Superconductivity and charge-density-wave in the Holstein model 
on the Penrose Lattice}
\author{Lu Liu}
\affiliation{School of Physics, Beijing Institute of Technology, Beijing 100081, China}
\author{Zi-Xiang Li}
\email{zixiangli@iphy.ac.cn}
\affiliation{Beijing National Laboratory for Condensed Matter Physics, Institute of Physics, Chinese Academy of Sciences, Beijing 100190, China}
\affiliation{University of Chinese Academy of Sciences, Beijing 100049, China}
\author{Fan Yang}
\email{yangfan_blg@bit.edu.cn}
\affiliation{School of Physics, Beijing Institute of Technology, Beijing 100081, China}

\begin{abstract}
The exotic quantum states emerging in the quasicrystal (QC) have attracted extensive interests because of various properties absent in the crystal. In this paper, we systematically study the Holstein model at half filling on a prototypical structure of QC, namely rhombic Penrose lattice, aiming at investigating the superconductivity (SC) and other intertwined ordering arising from the interplay between quasi-periodicity and electron-phonon ({\it e}-ph) interaction. 
Through unbiased sign-problem-free determinant quantum Monte Carlo simulations, we reveal the salient features of the ground state phase diagram. Distinct from the results on bipartite periodic lattices at half filling,  SC is dominant in a large parameter regime of the {\it e}-ph coupling (EPC) strength on the Penrose lattice. The strongest SC emerges in the intermediate EPC strength regime, where it coexists with the charge-density-wave (CDW). The CDW dominates the SC only in the sufficiently strong EPC regime. Our results also reveal pronounced pairing fluctuation above the transition temperature $T_c$ of the SC. The strong pairing and its fluctuation originate from the cooperative effects of the QC structure without translational symmetry and the macroscopically degenerate confined states at the Fermi energy, which uniquely exist on the Penrose lattice. Our unbiased numerical results suggest that the Penrose lattice is a potential platform to realize strong SC pairing, providing a promising avenue to 
discover relatively high-$T_c$ SC predominantly induced by the EPC.
\end{abstract}

\maketitle


\textcolor{ZXBlue}{\it Introduction}--- 
The discovery of quasicrystal (QC) \cite{sch1} has spurred exciting experimental and theoretical progresses for several decades. QCs possess self-similarity and long-range orientational order but lack translational symmetry. Moreover, QCs exhibit rotational symmetries, such as five-folded or eight-folded ones, that are forbidden in conventional crystals. Such special structure results in numerous exotic physical properties distinct from those of crystals, which include novel electronic states \cite{inayoshi,rmpjagannathan,moon,socolar-fujiwara,roberts2020,sakai,hauck,sakai2022,sakai2022-2,sakai2022-3,lesser2022,keskiner2022,nagai,sakai2023,jagannathan2023,jagannathan2023-2,ciardi2023,keskiner,sponza2024,hecker2024,weiren,kim2024,hori2024,crosse2021,jummojeon2022}, unique topological phases \cite{bandres2016,kraus,huang2018,longhi2019,peng2021,jeon2022,cwang2022,bhola2022,ghadimi2023,rchen2023,yang2024,madsen,flicker,rchen2023,junmojeon2024}. Increasingly more attentions have been paid to the correlated systems on QCs \cite{jagannathan2007,jagannathan2012,wessel2003,shaginyan2013,thiem2015,andrade2015,otsuki2016,koga2017,miyazaki2020,koga2021,koga2022,ghosh2023,ionue,khansili} since the discovery of quantum critical behavior in the Au-Al-Yb QC \cite{qc-exp}. Particularly, the superconductivity (SC), a typical macroscopic quantum phenomenon, has been observed in the {\rm Al-Zn-Mg} QC~\cite{sc-exp} and more recently in the {\rm Ta-Te} QC~\cite{terashima2024,tokumoto2024}, along with the earlier results on ternary QCs \cite{wong1987,wagner1988} and crystalline approximates \cite{deguchi2015}, stimulating extensive experiment and theoretical investigation to understand the intriguing properties of the SC on the QCs \cite{grai,grai2,uri2023,sakai2017,jhou2018,autti2018,araujo1,sakai2019,takemori,nagai2020,shiino2021,khosravian2021,yubo2022,fukushima2023,fukushima2023-2,ybliu2023,yecao,hori2024-1,sandberg2024,kobialka2024,ghadimi,saito2020,ghadimi2021,hori2024-2,cunyuanjiang}.

\begin{figure}
\centering
\includegraphics[width=86mm]{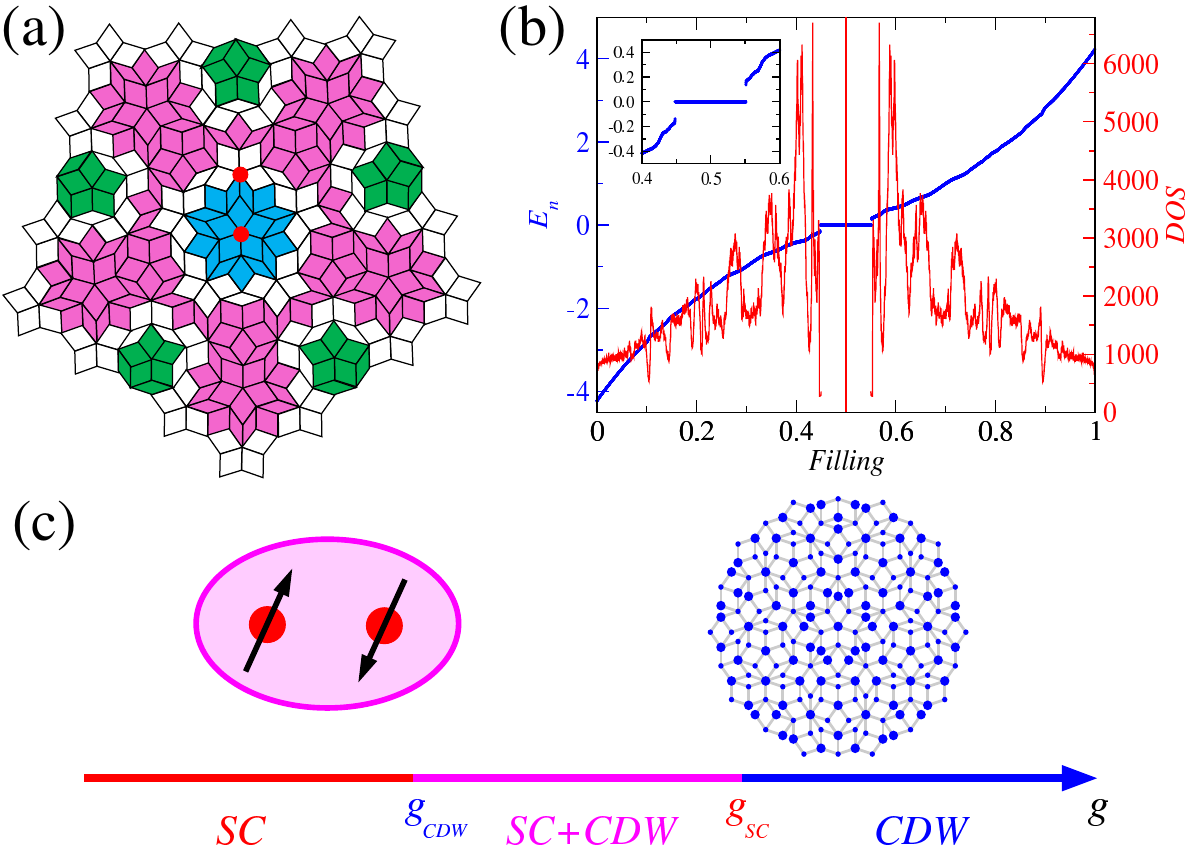}
\caption{(a) The different cluster regions (green, magenta, blue) and forbidden ladders (white) on the Penrose lattice. We have not shown the whole part of green cluster, which is connected to the edge of the largest forbidden ladder shown in this figure. The two red dots are the typical positions used in \Fig{spectrum}. (b) Single-particle energy spectrum and DOS on the Penrose lattice with $13926$ sites. Inset is a magnified graph for energy spectrum around half-filling. (c) Schematic phase diagram of the Holstein model on the Penrose lattice. When the EPC is weak $g<g_{\rm CDW}$, the ground state is SC. When $g>g_{\rm SC}$, the ground state is CDW state. There is an SC-CDW coexistence phase between  $g_{\rm CDW}$ and $g_{\rm SC}$.}
 \label{penrose}
 \end{figure}

The reported Bardeen-Cooper-Schrieffer (BCS) behaviors~\cite{sc-exp,terashima2024,tokumoto2024} for the SC observed in these QCs suggest the relevance of the electron-phonon ({\it e}-ph) interaction in the pairing mechanism. In the context of {\it e}-ph coupling (EPC), previous studies of the SC on the QCs are focused on the simplified attractive Hubbard model with the treatment of mean-field (MF) calculations~\cite{araujo1,sakai2017,sakai2019,nagai2020,takemori,yubo2022}. The attractive Hubbard model, as an effective interacting model accounting for the EPC, does not capture the retardation effect in the electronic interaction mediated by phonon fields, which substantially affects the ordering tendencies of SC and other instabilities. Particularly, in the regime of small phonon frequencies, relevant to most realistic materials, the electronic instabilities such as the charge-density-wave (CDW) and bipolaron are driven by the EPC, tremendously suppressing the transition temperature ($T_c$) of the SC \cite{kivelson2018,lee2019,AlexandreReview}. The competition between the SC and other electronic instabilities arising from EPC on the QC is not considered in previous studies. Moreover, the superconducting fluctuation beyond the MF calculations generally influences the phase coherence of pairing~\cite{kivelson1995}. This aspect might be more serious on the QCs as the lack of translational symmetry might hinder the formation of phase coherence~\cite{ybliu2023}. Hence, the investigation on the {\it e}-ph interaction on the QC through an unbiased approach incorporating all these crucial ingredients is immensely desired, which will definitely lead a great leap to fathoming the properties of phonon mediated SC and other orderings such as the CDW on the QCs.

To address this critical issue, we perform unbiased determinant quantum Monte-Carlo (DQMC) simulation on the Holstein model~\cite{holstein}, a prototypical microscopic model characterizing the EPC, on a representative QC, the Penrose lattice \cite{pen-til}. This work is the first systematic investigation on the consequence of the EPC 
on the QC, particularly focusing on the interplay between the SC and CDW, through DQMC approach without 
uncontrolled approximations. We focus on the half filling, at which the Fermi level accommodates macroscopically degenerate confined states~\cite{kohmoto,arai}, leading to divergent density-of-state (DOS). The DQMC results reveal salient features of the phase diagram with varying EPC strength. 
Remarkably, SC is the dominant instability for weak EPC. The SC persists up to an intermediate EPC regime, coexisting with the CDW, where its $T_c$ is maximized. The CDW only dominates the SC for strong enough EPC. Such a phase diagram is distinct from those of Holstein models on the half-filled bipartite crystals, in which the SC is always suppressed, either by the enhanced CDW induced by the Fermi-surface nesting (FSN) or by the vanishing DOS at the Dirac points. On the Penrose QC, while the absence of FSN suppresses the CDW, the divergent DOS strongly enhances the pairing strength, providing a promising route to enhancing SC mediated by EPC. Moreover, an analysis on the local DOS (LDOS) reveals the pseudo-gap phenomenon above the $T_c$, caused by the localized feature of the confined states.

\textcolor{ZXBlue}{\it Model and method}--- The Penrose lattice is a prototypical 2D QC. It is a bipartite lattice constructed from two types of rhombuses, as shown in Fig.\ref{penrose}(a). One important feature of the Penrose lattice is the macroscopically degenerate single-particle states~\cite{kohmoto,arai}. These states are dubbed as {\it confined states} and their wave functions are strictly confined in a finite domain region on the lattice. 
Several domains will overlap and cover a larger part dubbed as a cluster, which is the colored region shown in \Fig{penrose}(a). Adjacent clusters are separated by a strip region dubbed as the forbidden ladder \cite{arai,koga2017}. For details, see Supplemental Material (SM) \cite{supply}  
More intriguingly, the energies of the macroscopically degenerate confined states are exactly zero, rendering the infinite DOS at $E=0$, as shown in \Fig{penrose}(b). In the following, we scrutinize the interplay between these unique electronic structures and EPC on the Penrose lattice.

\begin{figure}
\centering
\includegraphics[width=85mm]{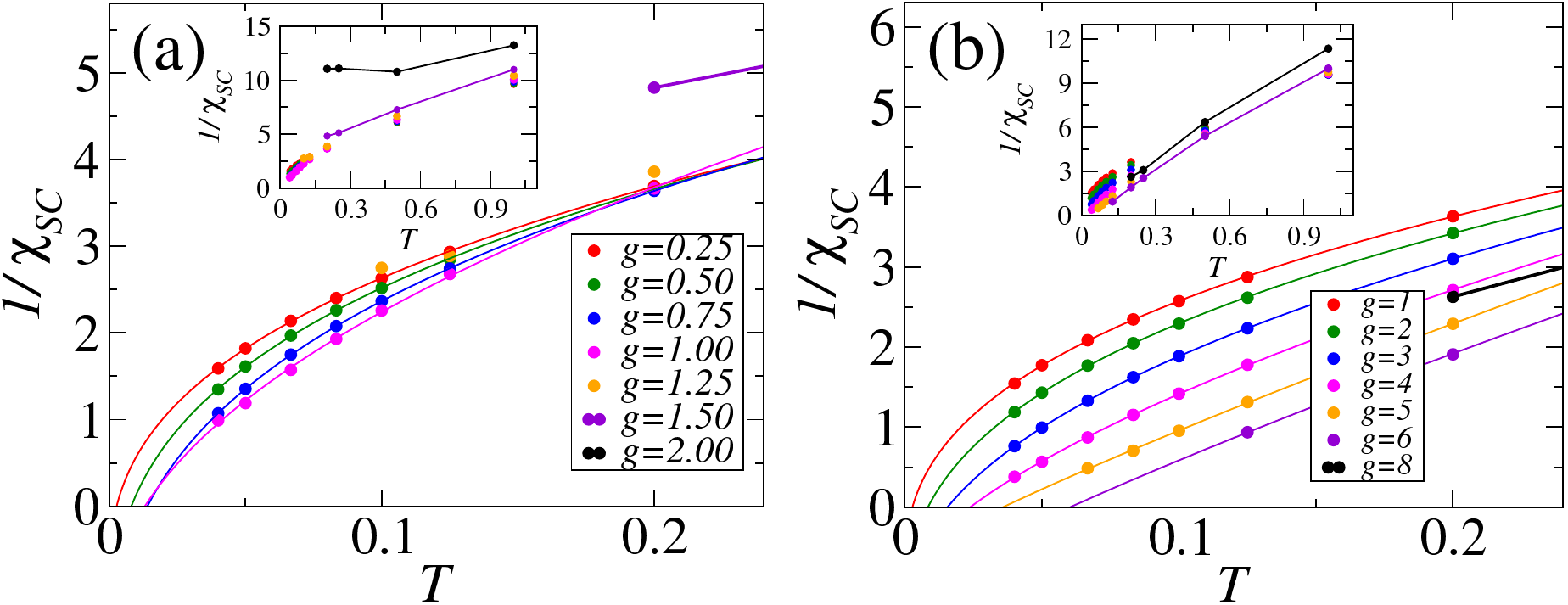}
\caption{The inverse of SC susceptibilities versus temperature $T$ for different EPC strength $g$ with $N=201$. The phonon frequency is (a). $\Omega=1$ and (b). $\Omega=4$.
The solid lines in (a) and (b) are the power law fit of susceptibilities. The fit function is $\chi_{\rm SC}^{-1}(T)=a+bT^c$.
Insets in (a) and (b) present the whole range of temperatures simulated.} 
\label{sus-sc}
\end{figure}

To numerically study the effect of EPC, we consider a typical {\it e}-ph model, termed as Holstein model, with the Hamiltonian reading:
\be
&&H=-t\sum_{\la ij\ra\sigma}(c_{i\sigma}^\dagger c_{j\sigma}+h.c.-\mu\sum_{i\sigma}c_{i\sigma}^\dagger c_{i\sigma}\nonumber\\
&&+\sum_i\left(\frac{p_i^2}{2M}+\frac{1}{2}M\Omega^2x_i^2 \right)-g\sum_i(n_i-\frac{1}{2})x_i.
\label{Ham}
\ee
Here $c_{i\sigma}^\dagger(c_{i\sigma})$ is creation (annihilation) operator for an electron on site $i$ with spin $\sigma$, $\mu$ is the chemical potential and $n_i$ is the number operator
on site $i$. Electrons are locally coupled with dispersionless phonon field with strength $g$.
 $x_i$ and $p_i$ are the position and momentum operators of phonon with mass $M$ and frequency $\Omega$ on site $i$. 
 The first sum of $\la ij\ra$ represents all nearest neighbor pairs of sites.
 We consider the model on the Penrose lattice and the sum of $\la ij\ra$ represents all edges of the rhombuses. 
 We set $M=t=1$ as the units of mass and energy, respectively. In this paper, we study two different frequencies $\Omega=1$ and $\Omega=4$. Because the model in \Eq{Ham} respects particle-hole symmetry, we fix the system at half filling by tuning $\mu=0$. 

We implement the DQMC algorithm to simulate the finite-temperature properties of \Eq{Ham}. 
Due to the symmetry where up and down spins couple in an identical way to the phonon field,
the model 
is free of the notorious minus sign problem \cite{method,method1,assaadreview,lireview,Wu2005prb,Li2016prl}, enabling the numerically accurate simulations 
with large system sizes and low temperatures. The details of simulation are included in SM~\cite{supply}.

\begin{figure}
\centering
\includegraphics[width=85mm]{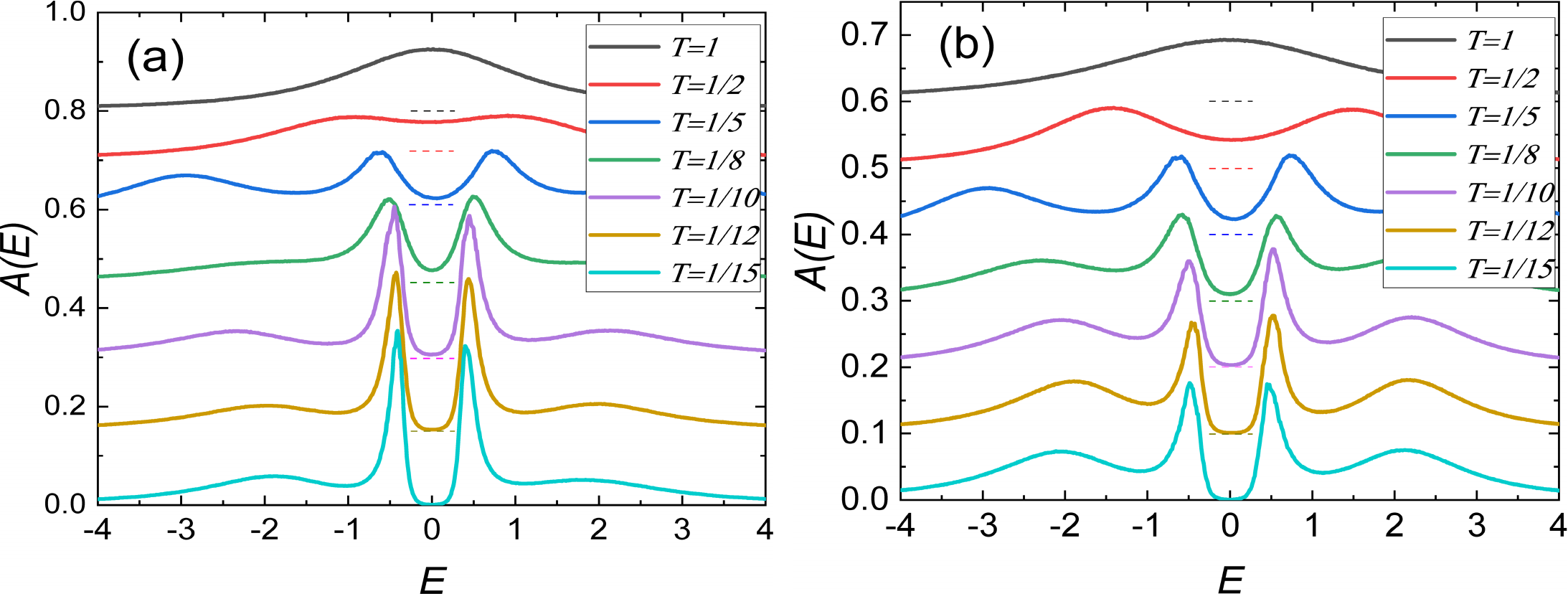}
\caption{The results of LDOS with varying temperature calculated at (a) the center of blue cluster and (b) the edge of forbidden ladder (shown in \Fig{penrose}(a)). The parameters are fixed as $\Omega = 4$ and $g=4$. The dash lines indicate the position of zero for each curve of LDOS. }
\label{spectrum}
\end{figure}

\textcolor{ZXBlue}{\it Superconductivity:} To characterize the SC ordering, we measure the SC susceptibility defined as: 
\be
\chi_{\rm SC}=\frac{1}{N}\int_0^\beta \la \Delta(\tau)\Delta^\dagger(0)\ra d\tau,
\ee
where $\beta$ is the inverse temperature and $\Delta$ is the order parameter of on-site SC defined as $\Delta=\sum_{i=1}^{N}\frac{1}{2}(c_{i\uparrow}^\dagger c_{i\dn}^\dagger +c_{i\dn}c_{i\up})$, where $N$ is the number of lattice sites.

We simulate the Holstein model on the Penrose lattice for six different site numbers, namely $N=31, 46, 76, 101, 141, 201$,  with open boundary condition. Here, we present the results for $N=201$. The results for other system sizes 
coincide with the results for $N=201$ qualitatively. We compute the susceptibility $\chi_{\rm SC}$ to detect the ordering tendency of the SC. We fix phonon frequency $\Omega = 1$ and $\Omega = 4$, respectively, and vary the strength of EPC $g$. With lowering temperature, $\chi_{\rm SC}$ exhibits the divergent behavior as long as the order is long-range (or quasi-long-range). Hence, to reveal the presence of SC more explicitly, we present the inverse of $\chi_{\rm SC}$ as shown in \Fig{sus-sc}. For both cases of $\Omega=1$ and $\Omega = 4$, $\chi_{\rm SC}$ increases with $g$ in the weakly coupling regime. For $\Omega=1$, the power-law fit of $1/\chi_{\rm SC}$ versus temperature $T$ indicates the divergence of $\chi_{\rm SC}$ at sufficiently low temperature when $g <1.25$, suggesting the presence of SC at ground state. With further increasing $g$, $\chi_{\rm SC}$ decreases. For instance, $\chi_{\rm SC}$ saturates at low temperature for $g=2$, indicating the absence of SC.  

With increasing phonon frequency from $\Omega=1$ to $\Omega=4$, the SC is enhanced as evidenced by the increase of $\chi_{\rm SC}$. For $\Omega=4$, the SC is enhanced with increasing EPC strength in a large parameter regime, 
being strongest at $g = 6$. The power-law fit of $1/\chi_{\rm SC}$ confirms the existence of SC at the ground state for $g \le 6$. For $\Omega=4$, the $\chi_{\rm SC}$ displays the feature of divergence at a relatively high temperature. For instance, for $g=6$, the fitting of $1/\chi_{\rm SC}$ showcases the divergence of $\chi_{\rm SC}$ occurring at $T \simeq 0.06$, implying the possibility of relatively high $T_c$ of SC. 
To confirm the robustness of the fitting results, we adopted another scaling function to fit $1/\chi_{\rm SC}$ according to the Berezinskii–Kosterlitz–Thouless scaling, which renders the approximately same results of the transition temperature, see the SM~\cite{supply}
As the EPC is further increased, the SC is suppressed as revealed by the decrease of $\chi_{\rm SC}$. Note that the fitted $T_c$ is subject to the finite size effect, which is expected to be slightly lower than the real $T_c$, see the SM~\cite{supply}.

To further investigate the SC properties on the Penrose lattice, we evaluate the LDOS, an observable measurable in the scanning tunnelling microscope, through stochastic analytical continuation of single-particle time-dependent Green's function~\cite{method2} (see \cite{supply} for details). We choose parameter in the regime where the SC is strong by fixing $\Omega=4$ and $g=4.0$, and calculate the LDOS at two typical positions on the Penrose lattice. At sufficiently low temperature, such as $T = 1/15$, the spectral gap is fully opened and two sharp coherence peaks with particle-hole symmetry appear, suggesting the emergence of SC. As depicted in \Fig{spectrum} (a) and (b), while the sizes of the gaps in the cluster center and those in the forbidden ladder are approximately the same, the coherence peaks in the latter are slightly broader than those in the former.  With increasing temperature to $T \geq 1/10$, the peaks evidently broaden but remain well-defined. Intriguingly, the feature of spectral gap, as revealed by the suppression of the spectral function at the Fermi energy, persists up to a temperature significantly higher than the $T_c$ of SC determined by the divergence of $\chi_{\rm SC}$, suggesting the existence of pseudo-gap behavior above the $T_c$. Since the spectral peaks are particle-hole symmetric, the pairing of electrons is a possible origin of the (pseudo) gaps above $T_c$\cite{kivelson1995}. Despite the strong pairing above $T_c$, the phase coherence is not achieved due to the localized feature of the confined states. Consequently, our results indicate the strong SC pairing is formed on the Penrose lattice induced by EPC, rendering a wide pseudo-gap regime induced by phase fluctuation above the $T_c$ of SC.

\begin{figure}
 \centering
 \includegraphics[width=85.mm]{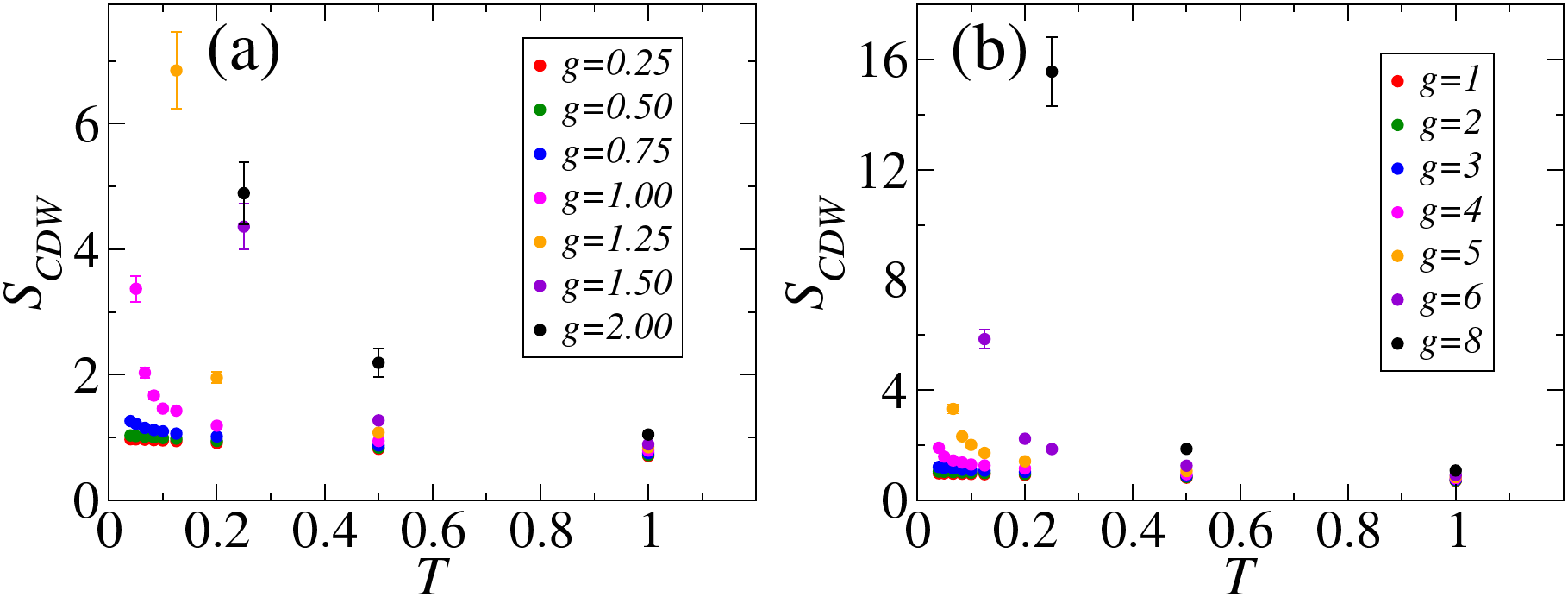}
 \caption{The CDW structure factor versus temperature $T$ for different EPC strengths $g$ with $N=201$. The phonon frequencies are (a) $\Omega=1$ and (b) $\Omega=4$ respectively.}
 \label{struc-cdw-new}
 \end{figure}

\textcolor{ZXBlue}{\it Charge-density-wave:} 
In addition to the SC, the CDW is usually another prevalent instability triggered by EPC, particularly in the strong EPC regime. For half-filled Holstein model on bipartite lattices, the CDW is generally favorable at finite phonon frequency. For example, the dominant instability is CDW with momentum $(\pi,\pi)$ on the square lattice, which is attributed to the FSN~\cite{bardley2021,hohenadler2019}. On the Penrose lattice, although the lattice momentum is not a good quantum number, the CDW order parameter is well defined as the difference of particle number between the two sublattices due to its bipartite nature.
The CDW order parameter and the associated structure factor are explicitly expressed as:
\be
O_{\rm CDW}&=&\frac{N}{2}\left(\frac{1}{N_A}\sum_{i\in A}^{N_A}n_i-\frac{1}{N_B}\sum_{j\in B}^{N_B}n_j\right), \\
S_{\rm CDW}&=&\frac{1}{N}O_{\rm CDW}^2.
\label{eq-cdw}
\ee
Here, $A/B$ labels the sublattices on the Penrose lattice, and $N_A/N_B$ denotes the corresponding site numbers. 

To detect the CDW order, we calculate its structure factor $S_{\rm CDW}$ defined in Eq.(\ref{eq-cdw}). 
\Fig{struc-cdw-new} (a) and (b) present the results of $\Omega=1$ and $\Omega=4$, respectively. The results reveal that when $g$ is small, $S_{\rm CDW}$ changes little as the temperature decreases, implying the absence of CDW long-range order at the ground state. However, when EPC is strong, the CDW is significantly enhanced. At $\Omega=1$,  $S_{\rm CDW}$ increases rapidly with lowering temperature when $g\ge 1.0$. The divergent behavior suggests the emergence of CDW long-range order at sufficiently low temperature. With the enhancement of CDW, the SC is suppressed. Nevertheless, the SC and CDW coexist at several values of $g$ in the intermediate coupling regime such as $g=1.0$. The increase of phonon frequency weakens the CDW and enhances the SC. At $\Omega=4$, the CDW is short-ranged in a larger coupling regime $g \le 4.0$, compared with the results for $\Omega=1$. When $g \ge 4.0$, $S_{\rm CDW}$ increases significantly and exhibits divergent behavior with lowering temperature, suggesting the existence of long-range order. For $\Omega=4$, the coexistent regime of SC and CDW is also larger, which is $4.0\le g \le 6.0$, compared with the case for $\Omega=1$.

\begin{figure} 
  \centering
  \includegraphics[width=85mm]{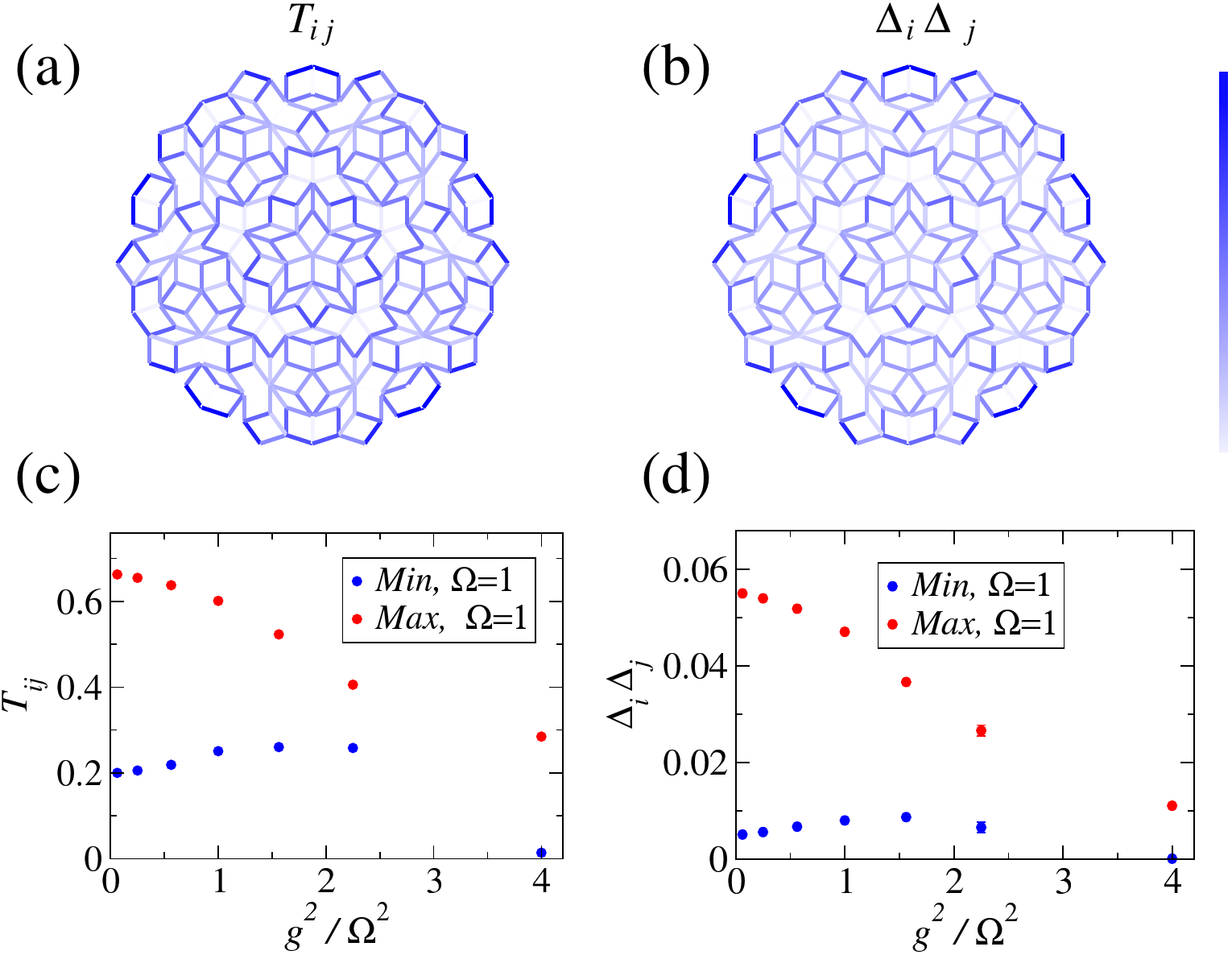}
\caption{The spatial pattern of (a) nearest neighbor hopping $T_{ij}$ and (b) nearest neighbor SC correlations $\Delta_i\Delta_j$ for $g=1$ with $\Omega=1$. 
The color bar on the right codes the value from the minimum to the maximum. (c), (d) show all the minimum and maximum values of $T_{ij}$ and $\Delta_i\Delta_j$ for different EPC strengths with $\Omega=1$ respectively. The minimum values of $T_{ij}$ and $\Delta_i\Delta_j$ are both on the forbidden ladders.}
 \label{pattern}
  \end{figure}

\textcolor{ZXBlue}{\it Enhancement of pairing on Penrose lattice:} The ground state phase diagram is schematically presented in \Fig{penrose} (c). The ground state is SC in the weak EPC regime, which is enhanced with increasing the coupling strength $g$. In the presence of strong EPC, the CDW is dominant which suppresses the SC. There is a regime of $g$ in which SC and CDW coexists. With increasing phonon frequency, this regime is notably enlarged. 

This phase diagram is qualitatively different from the ones for the Holstein models on bipartite periodic lattices, wherein the related band structures either possess FSN or feature Dirac points at the Fermi energy at half-filling. In the situation of FSN, typically on the square lattice, the ground state is generally CDW in the whole coupling regime for the finite phonon frequency\cite{bardley2021,hohenadler2019}.  For the systems featuring Dirac points,  the Dirac semi-metal phase is robust against weak EPC, and the CDW emerges in the strong EPC regime~\cite{yx-zhang,chuangchen}. In contrast, the FSN is absent on the Penrose QC due to lack of well-defined lattice momentum, resulting in the absence of CDW long-range order at weak coupling regime. 

Without CDW competing with SC at weak coupling regime, the SC on the Penrose lattice can be analyzed in the framework of the BCS theory. In the BCS theory, it is known that an infinitesimal attractive interaction induced by the EPC can lead to the Cooper instability. Such a Cooper instability for the Penrose lattice was analytically proven in Ref.~\cite{yubo2022}. Therefore, it is natural that the ground state is SC for the weak and intermediate EPC regimes. Additionally, in the BCS theory, the pairing strength enhances with the DOS. The DOS on the half-filled Penrose lattice shown in \Fig{penrose}(b) indicates a pronounced zero-energy peak. This divergent DOS strongly enhances the pairing amplitude under the EPC, which is also confirmed by the spectral results.

\textcolor{ZXBlue}{\it The effect of forbidden ladders:} As aforementioned, on the Penrose lattice, a large number of zero-energy states are confined in a region separated by the forbidden ladders, dubbed as confined states, giving rise to the divergent zero-energy DOS. Although the pairing amplitude is enhanced by the divergent DOS, the nature of confinement is expected to suppress the phase coherence of the pairing. To characterize such effect, we compute the expectation values of the hopping term $T_{ij}=\frac{1}{2}(c_i^\dagger c_j+c_j^\dagger c_i))$ and pair correlation $\Delta_i \Delta_j$ on each nearest neighbor pairs. The results for $\Omega=1$ and $g=1$ are presented in \Fig{pattern}(a) and (b) (see \cite{supply} for $\Omega=4$ results). 
The values of $T_{ij}$ and $\Delta_i \Delta_j$ are significantly weakened in the forbidden ladders. It suggests the suppression of SC pairing coherence in this region. With increasing EPC strength, the amplitudes of $T_{ij}$ and $\Delta_i \Delta_j$ in the forbidden ladders, as dictated in \Fig{pattern}(c) and (d), increase in the weak EPC regime and exhibit maximum values in the intermediate EPC regime, in alignment with the results of $\chi_{\rm SC}$ shown in \Fig{sus-sc}. The results suggest that the property of confined states separated by the forbidden ladder is a crucial ingredient affecting the SC on the Penrose lattice. The nature of localization for the confined states is suppressed in the presence of intermediate EPC, yielding the enhancement of the SC phase coherence.

It is intriguing that the physics of the superconductor-insulator-superconductor (SIS) junction could emerge on the Penrose lattice in the presence of SC due to the existence of forbidden ladders. The confined states form SC pairing, and every cluster can be thought of as a pairing island. The different pairing islands separated by the forbidden ladders are connected through the Josephson tunneling of Cooper pairs, which resembles the physics of SIS junctions. With increasing EPC in the weakly coupling regime, the phase coherence of the pairing in different clusters is enhanced, resulting in the enhancement of the SC long-range order. 


\textcolor{ZXBlue}{\it Discussions and Conclusions:}

In conclusion, our unbiased DQMC simulation of the Holstein model on the Penrose QC reveals that the SC is dominant in the weak and intermediate regime of EPC. Particularly, the strongest SC emerges
at the intermediate coupling regime, wherein it coexists
with CDW. Such a salient phase
diagram is brought about by the unique electronic structure of the Penrose lattice, which thus provides a promising platform to realize relatively high-Tc SC driven by the EPC.

We acknowledge that the combined challenges of rapidly diverging autocorrelation times in the low-temperature Holstein model and the intrinsic difficulty of resolving Kosterlitz-Thouless transitions, even in classical models, necessarily restrict the accuracy of transition temperature determination in this study. Despite these challenges, our detailed analysis of boundary, size, and shape effects ensures the robustness of our conclusions.




The Holstein model is the simplest one to describe the EPC, 
and we have further neglected the site-dependent coupling parameters~\cite{jlos} for simplicity
. More sophisticated model such as the SSH model~\cite{ssh1,ssh2,ssh4,ssh3,ssh5,ssh6,ssh7} is also amendable to the sign-free DQMC, albeit with more technical difficulties. The SSH model can capture more features of the lattice vibration in the QC, including the presence of the low-energy phason mode~\cite{phason1,phason2}, and thus it provides a viable platform to study the role of the phason mode in mediating the SC. We leave this topic for the future study.

~~~~

~~~~
\noindent{{\bf Acknowledgement}}

~~~~~~
The DQMC simulations are carried out with the ALF Library\cite{method1}. L.L. is supported by the National Natural Science Foundation of China under the Grant No. 12304171 and Beijing Institute of Technology Research Fund Program for Young Scholars. Z.L. is supported by the National Natural Science Foundation of China under the Grant No. 12347107. F.Y. is supported by the National Natural Science Foundation of China under the Grant Nos. 12234016, 12074031.

~~~~~


~~~~~


~~~~~~



\end{document}